\documentclass[twocolumn,showpacs,preprintnumbers,amsmath,amssymb]{revtex4}

\usepackage{graphicx}
\usepackage{dcolumn}
\usepackage{bm}
\usepackage{cases}
\usepackage{color}

\newcommand{\tb}{\textbf}

\begin{document}


\title{Pauli equation for a charged spin particle on a curved surface in an electric and magnetic field}
\author{Yong-Long Wang$^{1,2,}$}
 \email{wylong322@163.com}
\author{Long Du$^{1}$}
\author{Chang-Tan Xu$^{2}$}
\author{Xiao-Jun Liu$^{1,}$}
\email{liuxiaojun@nju.edu.cn}
\author{Hong-Shi Zong$^{3,4,5,}$}
\email{zonghs@nju.edu.cn}
\address{$^{1}$ Key Laboratory of Modern Acoustics, MOE, Institute of Acoustics, and Department of Physics, Nanjing University, Nanjing 210093, People's Republic of China}
\address{$^{2}$ Department of Physics, School of Science, Linyi University, Linyi 276005, People's Republic of China}
\address{$^{3}$ Department of Physics, Nanjing University, Nanjing 210093, People's Republic of China}
\address{$^{4}$ Joint Center for Particle, Nuclear Physics and Cosmology, Nanjing 210093, People's Republic of China}
\address{$^{5}$ State Key Laboratory of Theoretical Physics, Institute of Theoretical Physics, CAS, Beijing 100190, People's Republic of China}

\date{\today}

\begin{abstract}
We derive the Pauli equation for a charged spin particle confined to move on a spatially curved surface $\mathcal{S}$ in an electromagnetic field. Using the thin-layer quantization scheme to constrain the particle on $\mathcal{S}$, and in the transformed spinor representations, we obtain the well-known geometric potential $V_g$ and the presence of $e^{-i\varphi}$, which can generate additive spin connection geometric potentials by the curvilinear coordinate derivatives, and we find that the two fundamental evidences in the literature [Giulio Ferrari and Giampaolo Cuoghi, Phys. Rev. Lett. 100, 230403 (2008).] are still valid in the present system without source current perpendicular to $\mathcal{S}$. Finally, we apply the surface Pauli equation to spherical, cylindrical, and toroidal surfaces, in which we obtain expectantly the geometric potentials and new spin connection geometric potentials, and find that only the normal Pauli matrix appears in these equations.
\end{abstract}

\maketitle

\section{Introduction}
\indent The advent and development of nanostructure technology has renewed interest in an old subject, the dynamics constrained on a curved surface, for theoretical physicists. With the rapid development of the constructive nanotechnique and nanodevices \cite{Ahn2006, Ko2010}, the presence of various novel low-dimensional functional materials supplies diverse alternative experiments to test the developing theories. It is well known that the geometry experiences the motion of classical and quantum spatially limited particles in different ways. The curvature-induced geometric potential in nanostructures was discussed \cite{Encinosa1998, Ortix2011}, and some experiments gave some results to support that the curvature of a surface affects the dynamics of the system constrained on the surface \cite{Szameit2010, Onoe2012, Garcia2013}. The development of quantum theories for the dynamics of spatially reduced particles has become more and more important and urgent. The problem of a particle confined on a spatially curved surface has been considered for more than 50 years \cite{Dewitt1957, HJensen1971, Costa1981}, but up to the present it is still unsolved completely. The thin-layer quantization scheme \cite{Dewitt1957, HJensen1971, Costa1981, Costa1982, Encinosa2006, Ferrari2008, BJensen2009, BJensen2010, Ortix2011}, the generalized Dirac canonical quantization approach \cite{Liu2011, Liu2013, Liu2014}, and the path integral quantization formalism \cite{Matsutani1993, BJensen1993} are being developed to quantize the reduced particles. During the developments of these formalisms, the thin-layer quantization scheme was extended to describe $n$ noninteracting particles \cite{Costa1982}, and a spinless charged particle in an electric and magnetic field \cite{Encinosa2006, Ferrari2008, BJensen2009, BJensen2010, Ortix2011} by the surface Schr\"{o}dinger equation.\\
\indent The thin-layer quantization scheme is semiclassical. For a dimensionally reduced particle, its motion is described by the Schr\"{o}dinger equation, which is quantum, but the process constrained on a spatially curved surface is classical. The confining process is fulfilled by introducing a potential $V_{\lambda}(q_3)$, which has the limit $\lim_{q_3\rightarrow 0}V_{\lambda}=0$ and satisfies the condition $V_{\lambda}=\infty\quad (q_3\ne 0)$.\\
\indent We explore the presence of spin for a nonrelativistically charged particle constrained on a curved surface in the thin-layer quantization scheme. In the present paper, we derive the Pauli equation for the reduced spin system. This paper is organized as follows. In Sec. II, we derive the Pauli equation for a charged spin particle constrained on a spatially curved surface $\mathcal{S}$ and discuss the fundamental evidences given by Ferrari and Cuoghi \cite{Ferrari2008, Ortix2011} to decouple the Pauli equation into a normal component and a surface component. In Sec. III, we apply the surface Pauli equation to three examples, spherical, cylindrical, and toroidal surfaces, and obtain expectantly the well-known geometric potentials and additive spin connection geometric potentials, which are generated by the connection between spin and surface. In the three special examples, just the normal Pauli matrix appears in these surface Pauli equations. In Sec. IV, conclusions appear.
\section{Pauli Equation}
\begin{figure}[htbp]
  \centering
  \includegraphics[scale=0.39]{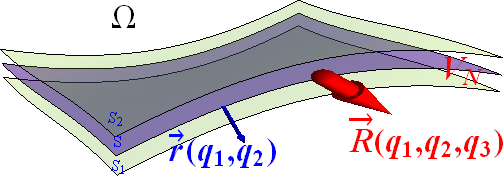}\\
  \caption{\footnotesize The surface $S$ and two auxiliary surfaces $S_1$ and $S_2$. The two surfaces close a subspace $V_N$ embedded in the ordinary 3D space $\Omega$. On the surface $S$, the point is described by $\vec{r}(q_1,q_2)$. In the subspace $V_N$, the point is described by $\vec{R}(q_1,q_2,q_3)$.}\label{Plane}
\end{figure}

\indent In the usual way, a common three-dimensional (3D) space $\Omega$ is described by three basis vectors $\vec{e}_1$, $\vec{e}_2$ and $\vec{e}_3$, which are perpendicular to each other. The portion of the immediate neighborhood of a surface $\mathcal{S}$, which is embedded in $\Omega$, is denoted by $V_N$ which is described by two basis vectors $\vec{u}_1$ and $\vec{u}_2$, which are locally parallel to the tangent plane of $\mathcal{S}$ at an arbitrarily fixed point, and a basis vector $\vec{n}$, which is normal to $\mathcal{S}$. Following the parametrization of da Costa \cite{Costa1981}, we employ equations $\vec{r}(q_1,q_2)$ to parametrize $\mathcal{S}$. The subspace $V_N$ sketched in Fig. \ref{Plane} may be parametrized by
\begin{equation}\label{Portion 1}
\vec{R}(q_1,q_2,q_3)=\vec{r}(q_1,q_2)+q_3\vec{n}(q_1,q_2),
\end{equation}
where $\vec{n}(q_1,q_2)$ is the normal basis vector depending only on $q_1$ and $q_2$. The total derivative of $\vec{R}(q_1,q_2,q_3)$ is
\begin{equation}\label{Portion 2}
d\vec{R}(q_1,q_2,q_3)=\vec{u}^1dq_1+\vec{u}^2dq_2+\vec{n}dq_3,
\end{equation}
where
\begin{equation}\label{Portion 3}
\vec{u}^1=\frac{\partial\vec{R}}{\partial q_1}=\frac{\partial\vec{r}}{\partial q_1}+q_3\frac{\partial\vec{n}}{\partial q_1},
\vec{u}^2=\frac{\partial\vec{R}}{\partial q_2}=\frac{\partial\vec{r}}{\partial q_2}+q_3\frac{\partial\vec{n}}{\partial q_2}
\end{equation}
and $\vec{n}$ are together called curvilinear coordinate system basis vectors.

\indent In $V_N$, and with the three basis vectors $\vec{u}_1$, $\vec{u}_2$, and $\vec{n}$, the covariant components of the metric tensor read
\begin{equation}\label{Metric Tensor 1}
G_{ij}=\frac{\partial\vec{R}}{\partial q^i}\cdot \frac{\partial\vec{R}}{\partial q^j}=G_{ji},\quad i,j=1,2,3.
\end{equation}
For the sake of clarity, we introduce the indices $a, b=1, 2$ to indicate the surface parameters. On $\mathcal{S}$, the covariant components of the metric tensor read
\begin{equation}\label{Metric Tensor 2}
g_{ab}=\frac{\partial\vec{r}}{\partial q^a}\cdot\frac{\partial\vec{r}}{\partial q^b}.
\end{equation}
As usual $G^{ij}$ and $g^{ab}$ denote the reciprocals of $G_{ij}$ and $g_{ab}$, respectively, and $G=\det(G_{ij})$, $g=\det(g_{ab})$. From Eq. \eqref{Portion 2}, it is shown that the derivatives of $\vec{n}(q_1,q_2)$ lie in the tangent plane of $\mathcal{S}$ at the point fixed by ($q_1,q_2$). And then the relations among $G_{ij}$ and $g_{ij}$ can be expressed as
\begin{equation}\label{Metric Tensor 3}
\begin{split}
&G_{ab}=g_{ab}+(\alpha g+g^T\alpha^T)_{ab}q_3+(\alpha g\alpha^T)_{ab}q_3^2,\\
&G_{a3}=G_{3a}=0,\quad G_{33}=1,
\end{split}
\end{equation}
through the Weingarten curvature matrix, whose elements read
\begin{equation}\label{Metric Tensor 4}
\alpha_a^b=-h_{ac}g^{cb},\quad h_{ac}=-\frac{\partial\vec{n}}{\partial q^a}\cdot\frac{\partial\vec{r}}{\partial q^c}.
\end{equation}
From Eqs. \eqref{Metric Tensor 3} and \eqref{Metric Tensor 4}, it follows that the relation between $G$ and $g$ satisfies the following expression
\begin{equation}\label{Metric Tensor 5}
G=f^2g,
\end{equation}
with
\begin{equation}
f=1+\mathrm{Tr}(\alpha)q_3+\det(\alpha)q_3^2.
\end{equation}
\indent In the presence of an externally applied electromagnetic field described by the potential vector $\vec{A}$ and the scalar electric potential $\phi_e$, we consider a spin particle with mass $m$ and charge $-e$, confined on a two-dimensional (2D) surface by introducing the squeezing potential $V_{\lambda}(q_3)$ \cite{HJensen1971,Costa1981}. For this system, the dynamics can be described by the Pauli equation
\begin{equation}\label{Pauli Equation 1}
\{-\frac{\hbar^2}{2m}[\vec{\sigma}\cdot(\vec{\nabla}+\frac{ie}{\hbar} \vec{A})]^2-e\phi_e+V_{\lambda}(q_3)\}\psi=i\hbar\frac{\partial}{\partial t}\psi,
\end{equation}
where $\vec{\sigma}$ has three components being ordinary Pauli matrices, $\vec{\nabla}$ denotes an ordinary derivative vector, and the wave function
\begin{equation}\label{Wavefunction 1}
\psi=
\left (
\begin{array}{c}
\psi_+\\
\psi_-\\
\end{array}
\right )
\end{equation}
contains two components. Using the Pauli vector identities, $(\vec{\sigma}\cdot\vec{a}) (\vec{\sigma}\cdot\vec{b}) =(\vec{a}\cdot\vec{b})+i\vec{\sigma}\cdot(\vec{a}\times\vec{b})$, we can remove the Pauli matrices from the kinetic energy term to rewrite \eqref{Pauli Equation 1} as
\begin{equation}\label{Pauli Equation 2}
\begin{split}
&\{-\frac{\hbar^2}{2m}[(\vec{\nabla}+\frac{ie}{\hbar}\vec{A})^2- \frac{e}{\hbar}\vec{\sigma}\cdot(\vec{\nabla}\times\vec{A})]\\
&-e\phi_e+V_{\lambda}(q_3)\}\psi=i\hbar\frac{\partial}{\partial t}\psi.
\end{split}
\end{equation}
Introducing the definitions $D_0=\partial_t-\frac{i e}{\hbar}\phi_e$ and $D_i=\nabla_i+\frac{ie}{\hbar}A_i$, we can simplify Eq. \eqref{Pauli Equation 2} in the following form
\begin{equation}\label{Pauli Equation 3}
\begin{split}
i\hbar D_0\psi=& -\frac{\hbar^2}{2m}G^{ij}D_iD_j\psi\\ &+\frac{e\hbar}{2m\sqrt{G}} \varepsilon^{ijk}\sigma_i\partial_jA_k\psi
 +V_{\lambda}(q_3)\psi.
\end{split}
\end{equation}
Under the following gauge transformations,
\begin{equation}\label{Gauge Transformations 1}
\begin{cases}
A_i^{\prime}=A_i+\partial_i\gamma,\\
A_0^{\prime}=A_0+\partial_t\gamma,\\
\psi^{\prime}=\psi e^{-ie\gamma/\hbar}
\end{cases}
\end{equation}
with $\gamma$ being an arbitrary scalar function, the gauge invariance of Eq. \eqref{Pauli Equation 3} is easily demonstrated.\\
\indent For the sake of calculational simplicity, we continuously use the curvilinear coordinate system which is the same as the framework in Eq. \eqref{Portion 1}. In the specially moving coordinate frame, there are some mathematical variables and operators that need to be briefly reviewed. The covariant derivatives $\nabla_i$ are defined by $\nabla_iv^j=\partial_iv^j+\Gamma_{ik}^jv^k$, where $v^j$ are the contravariant components of the 3D vector potential $\vec{v}$, $\partial_i$ are the derivatives with respect to the spatial variables $q_i$, and $\Gamma_{ik}^j$ are the Christoffel symbols defined by $\Gamma_{ij}^k=\frac{1}{2}G^{kl}[\partial_jG_{li}+\partial_iG_{lj} -\partial_lG_{ij}]$ \cite{Ferrari2008,Ortix2011}. With $\Gamma_{ij}^i=\frac{\partial\ln\sqrt{G}}{\partial q^j}$, the covariant divergence reads $\nabla_iA^i=\frac{1}{\sqrt{G}}\frac{\partial(\sqrt{G}A^i)}{\partial q^i}$, and then $\nabla^2=\frac{1}{\sqrt{G}}\frac{\partial}{\partial q^i}(\sqrt{G}G^{ij}\frac{\partial \psi}{\partial q^j})$. The rotation of $\vec{A}$ is $\vec{\nabla}\times\vec{A}=\xi^i\vec{e}_i$ with $\xi^i=\frac{1}{\sqrt{G}}\varepsilon^{ijk}\nabla_jA_k$, where the repeating of indexes is summation. The Levi-Civita symbol $\varepsilon^{ijk}$ is $1$ when $(ijk)$ is $(123)$ with even operations, $-1$ with $(ijk)$ being $(123)$ operated odd times, and $0$ in otherwise.\\
\indent In the curvilinear coordinate system, in terms of the ordinary Pauli matrices
\begin{equation}\label{Pauli Matrix 1}
\sigma^1=
\left (
\begin{array}{cc}
0 & 1\\
1 & 0\\
\end{array}
\right ),
\sigma^2=
\left (
\begin{array}{cc}
0 & -i\\
i & 0\\
\end{array}
\right ),
\sigma^3=
\left (
\begin{array}{cc}
1 & 0\\
0 & -1\\
\end{array}
\right ),
\end{equation}
the induced matrices\cite{Dick2006} can be defined by
\begin{equation}\label{Pauli Matrix 2}
\Sigma^a(q)=\sum_{i=1}^3\sigma^i\partial_iq^a,\quad a=1,2,
\end{equation}
which satisfy the following commutation relations
\begin{equation}\label{Pauli Matrix 3}
[\Sigma^a(q),\Sigma^b(q)]=2i\varepsilon^{ijk}\sigma_i \partial_jq^a\partial_kq^b,
\end{equation}
and anticommutation relations
\begin{equation}\label{Pauli Matrix 4}
\{\Sigma^a(q),\Sigma^b(q)\}=2g^{ab}(q),\quad a,b=1,2,
\end{equation}
where $q$ denotes ($q_1, q_2, q_3$), $i,j,k=1,2,3$, and the repeating of indexes means summation. In this special case, the induced Pauli matrices with respect to the basis vectors $\vec{u}_1$ and $\vec{u}_2$ depend on the three matrices in Eqs. \eqref{Pauli Matrix 1}. In order to gauge away $\sigma^3$ from the right-hand side of Eq. \eqref{Pauli Matrix 2}, we have to perform a rotation $\mathcal{R}(q)$ to bring the tangent plane of $\mathcal{S}$ into the ($\vec{e}_1,\vec{e}_2$) plane. The tangent plane is spanned by $\vec{u}_1$ and $\vec{u}_2$. In terms of the two basis vectors, we can construct the Cartesian vectors \cite{Dick2006}
\begin{equation}\label{Basis Vectors 1}\nonumber
\begin{cases}
\vec{n}_1=\frac{\vec{u}_1}{|\vec{u}_1|},\\
\vec{n}_2=\frac{\vec{u}_2-(\vec{n}_1\cdot\vec{u}_2)\vec{n}_1} {\sqrt{\vec{u}^2_2-(\vec{n}_1\cdot\vec{u}_2)^2}},\\
\vec{n}_3=\vec{n}_1\times\vec{n}_2=\frac{\vec{u}_1\times\vec{u}_2} {\sqrt{\vec{u}_1^2\vec{u}_2^2-(\vec{u}_1\cdot\vec{u}_2)^2}},
\end{cases}
\end{equation}
where $\vec{n}_3$ is parallel to $\vec{n}$ in Eq. \eqref{Portion 1}. And then we define a rotation $\mathcal{R}(q)$ which satisfies the following expression:
\begin{equation}\label{Rotation Tansform 1}
\mathcal{R}(q)
\left (
\begin{array}{c}
\vec{n}_1\\
\vec{n}_2\\
\vec{n}_3
\end{array}
\right )=
\left (
\begin{array}{c}
\vec{e}_1\\
\vec{e}_2\\
\vec{e}_3
\end{array}
\right ).
\end{equation}
The corresponding spinor representation $\mathcal{U}(q)=\mathcal{U}(\mathcal{R}(q))$ can gauge away the $\sigma^3$ term from the right-hand side of Eq. \eqref{Pauli Matrix 2}, which reads
\begin{equation}\label{Pauli Matrix 5}
\sigma^a(q)=\mathcal{U}(q)\Sigma^a(q)\mathcal{U}^{-1}(q)=\sum_{i=1}^2 \sigma^i\partial_iq^a.
\end{equation}
It is easy to check that the transformed induced matrices $\sigma^a(q)$ still satisfy the commutation relations
\begin{equation}\label{Pauli Matrix 6}
[\sigma^a(q),\sigma^b(q)]=2i\varepsilon^{ijk}\sigma_i \partial_jq^a\partial_kq^b,
\end{equation}
and the anticommutation relations
\begin{equation}\label{Pauli Matrix 7}
\{\sigma^a(q),\sigma^b(q)\}=2g^{ab}(q).
\end{equation}
In the transformed spinor representation, the Pauli matrices in Eq. \eqref{Pauli Equation 2} should be two transformed induced Pauli matrices $\sigma^1$ and $\sigma^2$ defined by Eq. \eqref{Pauli Matrix 5}, and one ordinary Pauli matrix $\sigma^3$ in Eq. \eqref{Pauli Matrix 1}, and the wave function $\psi$ should be replaced by $\mathcal{U}^{-1}(q)\psi$.\\
\indent According to the previous concise discussions, for a generic 3D curvilinear coordinate system, the covariant Pauli equation can be obtained as
\begin{equation}\label{Pauli Equation 4}
\begin{split}
i\hbar D_0\psi^{\prime}
=&-\frac{1}{2m}\{\frac{\hbar^2}{\sqrt{G}} \partial_i(\sqrt{G}G^{ij} \partial_j\psi^{\prime})\\
&+\frac{ie\hbar}{\sqrt{G}}[\partial_i(\sqrt{G}G^{ij}A_j)]\psi^{\prime}\\
&+2ie\hbar G^{ij}A_i\partial_j\psi^{\prime}-e^2G^{ij}A_iA_j\psi^{\prime}\\ &-\frac{e\hbar}{\sqrt{G}}\varepsilon^{ijk}\sigma_i(\partial_jA_k) \psi^{\prime}\}+V_{\lambda}(q_3)\psi^{\prime},
\end{split}
\end{equation}
where $\psi^{\prime}=\mathcal{U}^{-1}(q)\psi$. It is clear that the previous Pauli equation can return to the Schr\"{o}dinger equation in Ref. \cite{Ferrari2008} if the spin is ignored.\\
\indent The rotation defined in Eq. \eqref{Rotation Tansform 1} can be accomplished through two rotations around two axes, respectively, which are different for different points ($q_1,q_2$) on $\mathcal{S}$. We can have
\begin{equation}\label{Rotation Transform 2}
\mathcal{U}(q)=e^{i\varphi(q_1,q_2,\sigma)},
\end{equation}
where $\sigma$ may be $\sigma^1$, $\sigma^2$, and $\sigma^3$. According to the well-known thin-layer quantization scheme \cite{HJensen1971, Costa1981, Costa1982, Encinosa2006, Ferrari2008, BJensen2009, BJensen2010, Ortix2011}, we try to decouple the wave function $\psi$ into surface and normal parts by introducing a new wave function $\chi$, which is $\chi=\chi_s\chi_n=\sqrt{f}\psi$. From the structure of the metric tensor \eqref{Metric Tensor 3}, it is straightforward to show the following limiting relations:
\begin{equation}\label{Limit 1}
\lim_{q_3\rightarrow 0}G^{ab}\Gamma_{ab}^c=g^{ab}\gamma_{ab}^c,\quad
\lim_{q_3\rightarrow 0}G^{ab}\Gamma_{ab}^3=-Tr(\alpha).
\end{equation}
With the limit $q_3\rightarrow 0$, the corresponding relations between the original wave function $\psi$ and the new wave function $\chi$ and their derivatives are
\begin{equation}\label{Limit 2}
\begin{cases}
\lim_{q_3\rightarrow 0}\psi=\chi,\\
\lim_{q_3\rightarrow 0}\partial_3\psi=\partial_3\chi-\frac{1}{2}\mathrm{Tr}(\alpha)\chi,\\
\lim_{q_3\rightarrow 0}\partial_3^2\psi=\partial_3^2\chi-\mathrm{Tr}(\alpha)\partial\chi +\frac{3}{4}[\mathrm{Tr}(\alpha)]^2-\det(\alpha)\chi.
\end{cases}
\end{equation}
Defining a new metric tensor $\tilde{G}$ in the form
\begin{equation}
\tilde{G}=
\left (
\begin{array}{ccc}
g_{11} & g_{12} & 0 \\
g_{21} & g_{22} & 0 \\
0 & 0 & 1 \\
\end{array}
\right ),
\end{equation}
and introducing $\tilde{\nabla}_i v_j=\partial_i v_j-\gamma_{ij}^kv_k$, and $\tilde{D}_i=\tilde{\nabla}_i+\frac{ie}{\hbar}A_i$,
we can rewrite the Pauli equation \eqref{Pauli Equation 4} as
\begin{equation}\label{Pauli Equation Add1}
\begin{split}
i\hbar D_0\chi^{\prime}=&-\frac{1}{2m}\tilde{G}^{ij}\tilde{D}_i\tilde{D}_j \chi^{\prime}\\
&+\frac{e\hbar}{2m\sqrt{\tilde{G}}}\varepsilon^{ijk} \sigma_i\partial_jA_k\chi^{\prime}+V_{\lambda}(q_3)\chi^{\prime},
\end{split}
\end{equation}
where $\tilde{G}=\det(\tilde{G}_{ij})$, and $\chi^{\prime}=e^{-i\varphi}\chi$. It is easy to check that the Pauli equation \eqref{Pauli Equation Add1} is invariant under the gauge transformations \eqref{Gauge Transformations 1}. We therefore obtain a mapping of the original metric tensor $G_{ij}$ into $\tilde{G}_{ij}$ preserving the gauge invariance. Using the previously discussed results, we expand the Pauli equation \eqref{Pauli Equation Add1} as
\begin{widetext}
\begin{equation}\label{Pauli Equation 5}
\begin{split}
i\hbar D_0e^{-i\varphi}\chi=&-\frac{\hbar^2}{2m\sqrt{g}} \partial_a[\sqrt{g}g^{ab}\partial_b\chi^{\prime}]
-\frac{ie\hbar}{m}g^{ab}A_a\partial_b\chi^{\prime}
-\frac{ie\hbar}{2m\sqrt{g}}[\partial_a(\sqrt{g}g^{ab}A_b)]\chi^{\prime}
+\frac{e^2}{2m}g^{ab}A_aA_b\chi^{\prime}\\
&+\frac{e\hbar}{2m\sqrt{g}}\varepsilon^{3ab}\sigma_3(\partial_aA_b) \chi^{\prime}
+\frac{e\hbar}{2m\sqrt{g}}\varepsilon^{ab3}\sigma_a(\partial_bA_3) \chi^{\prime}
-\frac{\hbar^2}{2m}\partial_3^2\chi^{\prime}
-\frac{ie\hbar}{2m}\partial_3A_3\chi^{\prime}
-\frac{ie\hbar}{m}A_3\partial_3\chi^{\prime}\\
&+\frac{e^2}{2m}A_3^2\chi^{\prime}
+\frac{e\hbar}{2m\sqrt{g}}\varepsilon^{a3b}\sigma_a(\partial_3A_b) \chi^{\prime}+V_g\chi^{\prime}
+V_{\lambda}(q_3)\chi^{\prime},
\end{split}
\end{equation}
\end{widetext}
where $V_g$ is the well-known geometric potential in the form
\begin{equation}\label{Geometric Potential 1}
V_g=-\frac{\hbar^2}{2m}(M^2-K),
\end{equation}
wherein $M=\frac{1}{2}Tr(\alpha)$ is the mean curvature and $K=\det(\alpha)$ is the Gaussian curvature. In Eq. \eqref{Pauli Equation 5}, the term of mixing $A_j$ and the curvature matrix $\alpha_a^b$ does not appear.  With the metric tensor $\tilde{G}_{ij}$, the Lorentz gauge reads
\begin{equation}\label{Lorentz Gauge 1}
\vec{\nabla}\cdot\vec{A}=g^{ab}\partial_aA_b+\partial_3A_3.
\end{equation}
One of the two fundamental evidences given in Ref.\cite{Ferrari2008} is still valid in the present paper. There is no coupling between the magnetic field and the curvature of the surface, independently of the shape of the surface, of the field $\vec{B}$, and of the gauge, but dependent on the metric tensor $\tilde{G}_{ij}$.\\
\indent In the previously discussed results, the electromagnetic field is externally applied, and the corresponding source currents are not mentioned. In the trivial case, the coupling of the electromagnetic field to the curvature of the surface is trivially decoupled. In a general case of the present reduced system containing source currents $\vec{J}$ in all space directions, the vanishing of the $J^3$ component is a necessary condition to remove the coupling of the electromagnetic field to the curvature of the surface from the Lorentz gauge and the dynamics simultaneously \cite{BJensen2009}. The condition is also necessary to decompose the dynamics into a surface component and a transverse component. These results open up some interesting avenues to research a 2D reduced system.\\
\indent Although the Pauli equation \eqref{Pauli Equation 5} is more complicated than the Schr\"{o}dinger equation in \cite{Ferrari2008}, there is also only one term $A_3(q_1,q_2,0)\partial_3\chi^{\prime}$ connecting the dynamics on $\mathcal{S}$ with the dynamics along $q_3$. The term including $\partial_3A_a$ vanishes on $\mathcal{S}$ for the surface components of $\vec{A}$ depending only on the two tangent variables $q_1$ and $q_2$. With the metric tensor $\tilde{G}_{ij}$ to choose the best suitable choice
\begin{equation}\label{Gauge Choice 1}
\gamma(q_1,q_2,q_3)=-\int_0^{q_3}{A_3(q_1,q_2,z)dz}
\end{equation}
for the scalar function in Eq. \eqref{Gauge Transformations 1}, and with $\partial_aA_3$ depending only on $q_1$ and $q_2$ at $q_3\rightarrow 0$, we can eliminate all terms containing $A_3$ or $\partial_3A_3$ from Eq. \eqref{Pauli Equation 5}. We decompose the Pauli equation \eqref{Pauli Equation 5} into a normal component
\begin{equation}\label{Pauli Equation 6}
i\hbar\partial_t\chi_n=-\frac{\hbar^2}{2m}\partial_3^2\chi_n +V_{\lambda}(q_3)\chi_n,
\end{equation}
and a surface component
\begin{equation}\label{Pauli Equation 7}
\begin{split}
i\hbar\partial_t(e^{-i\varphi}\chi_s)=&-\frac{1}{2m}\{\frac{\hbar^2}{\sqrt{g}} \partial_a\sqrt{g} g^{ab}\partial_b\\
&+2ie\hbar g^{ab}A_a\partial_b+\frac{ie\hbar}{\sqrt{g}}[\partial_a(\sqrt{g}g^{ab}A_b)]\\
&-e^2g^{ab}A_aA_b-\frac{e\hbar}{\sqrt{g}}\varepsilon^{ab}\sigma_3 (\partial_aA_b)\\
&-\frac{e\hbar}{\sqrt{g}}\varepsilon^{ab}\sigma_a(\partial_bA_3) \}\chi_s^{\prime}\\
&+V_g\chi_s^{\prime}-e\phi_e\chi_s^{\prime},
\end{split}
\end{equation}
where the wave function $\chi_s^{\prime}=e^{-i\varphi}\chi_s$, wherein $\chi_s$ is described by the ordinary spinor representation containing two components,
\begin{equation}\label{Wavefunction 2}
\chi_s=
\left (
\begin{array}{c}
\chi_s^+\\
\chi_s^-
\end{array}
\right ).
\end{equation}
Expression \eqref{Pauli Equation 6} is just a one-dimensional Pauli equation for a charged spin particle constrained on $\mathcal{S}$ by the normal potential $V_{\lambda}(q_3)$ and can be ignored in all future calculations. However, expression \eqref{Pauli Equation 7} is very interesting because of the presence of $V_g$ and $e^{-i\varphi}$ and it describes the dynamics of the spin particle with charge $-e$ bounded on $\mathcal{S}$ in an electromagnetic field. In this case, the actions of the curvilinear coordinate derivative terms on $e^{-i\varphi}$ can bring out some spin connection geometric potentials. Ignoring the spin of the discussed system, from expression \eqref{Pauli Equation 7} we can vanish the terms depending on the transformed induced and normal Pauli matrices and obtain the surface Schr\"{o}dinger equations in Ref. \cite{Ferrari2008}. It is worth noticing that the decoupling of the Pauli equation \eqref{Pauli Equation 5} is analytical without approximation, but with the vanishing of the source current perpendicular to the surface $\mathcal{S}$. With $\tilde{G}_{ij}$, the second fundamental evidence in Ref. \cite{Ferrari2008} is also valid in the present paper. Without the source current normal to $\mathcal{S}$, the dynamics on the surface and the normal dynamics can be separated.
\section{Pauli Equations on Spherical, Cylindrical and Toroidal Surfaces}
\begin{figure}[htbp]
\centering
\includegraphics[scale=0.37]{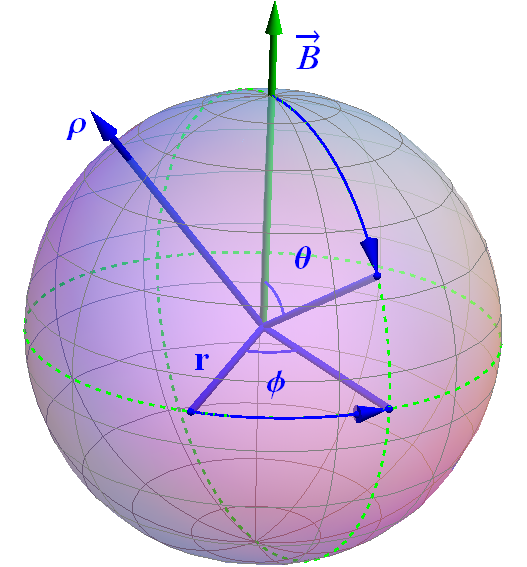}\\
\caption{\footnotesize A spherical surface of radius $r$ and its coordinate system $(\theta,\phi,\rho)$. The green arrow denotes the direction of the magnetic field $\vec{B}$, which is perpendicular to the plane at $\theta=\frac{\pi}{2}$.}\label{Sphere}
\end{figure}

\begin{figure}[htbp]
\centering
\includegraphics[scale=0.37]{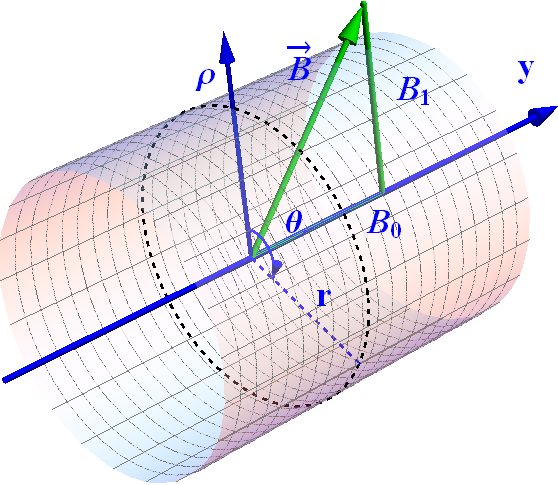}\\
\caption{\footnotesize A cylindrical surface of radius $r$ and its coordinate system $(\theta, y, \rho)$. The magnetic field $\vec{B}$ consists of $B_0$ parallel to the $y$ axis and $B_1$ perpendicular to the $y$ axis at $\theta=0$.}\label{Cylinder}
\end{figure}

\begin{figure}[htbp]
\centering
\includegraphics[scale=0.37]{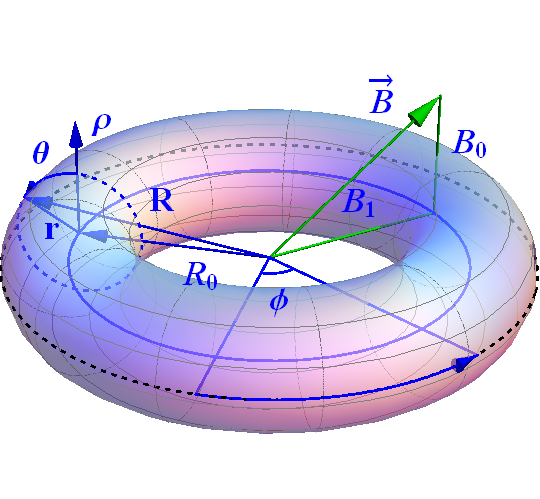}\\
\caption{\footnotesize A toroidal surface of radius $r$ and its coordinate system $(\theta, \phi, \rho)$. $R_0$ is the distance from the center of the tube to the center of the torus, $r$ is the radius of the tube, and $R$ is the distance from the center of the tube to one point which lies on the toroidal surface.  The magnetic field $\vec{B}$ can be decomposed into two parts: $B_1$ lies in the plane of $R_0$ and $\phi$, and $B_0$ is perpendicular to the circle defined by $R_0$ and $\phi$.}\label{Torus}
\end{figure}

\indent It is well known that the sphere and the cylinder are two typical nanostructures, and that the torus is a typical topological geometry. In the presence of a homogeneous magnetic field, we derive the surface Pauli equation for a charged spin particle constrained on the three surfaces. The spherical surface is the simplest one, and its extensive investigation is fullerene \cite{Onoe2012}. We first discuss it. In the spherical coordinate system $(\theta,\phi,\rho)$, a sphere with radius $r$ is put in a constant magnetic field $\vec{B}$, which is parallel to the polar axis, and is sketched in Fig. \ref{Sphere}.
\indent In spherical coordinates, on the sphere the induced Pauli matrices \eqref{Pauli Matrix 2} are
\begin{equation}\nonumber
\begin{split}
&\Sigma^{\theta}=\frac{1}{r}
\left (
\begin{array}{cc}
-\sin\theta & \cos\theta e^{-i\phi}\\
\cos\theta e^{i\phi} & \sin\theta
\end{array}
\right ),\\
&\Sigma^{\phi}=\frac{1}{r\sin\theta}
\left (
\begin{array}{cc}
0 & -ie^{-i\phi}\\
ie^{i\phi} & 0
\end{array}
\right ),\\
&\Sigma^{\rho}=
\left (
\begin{array}{cc}
\cos\theta & \sin\theta e^{-i\phi}\\
\sin\theta e^{i\phi} & -\cos\theta
\end{array}
\right ).
\end{split}
\end{equation}
On the sphere, for the transformation of the induced Pauli matrices, the spinor rotation is
\begin{equation}\nonumber
\begin{split}
\mathcal{U}(\theta,\phi)&=e^{\frac{i}{2}\theta\sigma_2 +\frac{i}{2}\phi\sigma_3}\\
&=
\left (
\begin{array}{cc}
\cos\frac{\theta}{2}e^{i\frac{\phi}{2}} & \sin\frac{\theta}{2}e^{-i\frac{\phi}{2}}\\
-\sin\frac{\theta}{2}e^{i\frac{\phi}{2}} & \cos\frac{\theta}{2}e^{-i\frac{\phi}{2}}
\end{array}
\right )
\end{split}
\end{equation}
and its inverse matrix is
\begin{equation}\nonumber
\mathcal{U}^{-1}(\theta,\phi)=
\left (
\begin{array}{cc}
\cos\frac{\theta}{2}e^{-i\frac{\phi}{2}} & -\sin\frac{\theta}{2}e^{-i\frac{\phi}{2}}\\
\sin\frac{\theta}{2}e^{i\frac{\phi}{2}} & \cos\frac{\theta}{2}e^{i\frac{\phi}{2}}
\end{array}
\right ).
\end{equation}
The transformed induced Pauli matrices on the sphere are
\begin{equation}\nonumber
\sigma^{\theta}=\frac{1}{r}\sigma^1,\quad
\sigma^{\phi}=\frac{1}{r\sin\theta}\sigma^2,\quad
\sigma^{\rho}=\sigma^3.
\end{equation}
According to the gauge condition \eqref{Gauge Choice 1}, the best suitable vector potential is $(A_{\theta},A_{\phi},A_{\rho})=(0,\frac{1}{2}Br^2\sin^2\theta,0)$. From Eq. \eqref{Pauli Equation 7}, the spherical surface dynamics for the charged spin particle can be described by
\begin{equation}\label{Pauli Equation 8}
\begin{split}
i\hbar\partial_t\chi_s^{\prime}=& -\frac{1}{2m}\{\frac{\hbar^2}{r^2}\partial_{\theta}^2 +\frac{\hbar^2\cos\theta} {r^2\sin\theta}\partial_{\theta} +\frac{\hbar^2}{r^2\sin^2\theta}\partial_{\phi}^2\\
&-(\frac{i\hbar^2}{r^2\sin^2\theta}\sigma_{\rho}-ie\hbar B)\partial_{\phi}\\
&-[\frac{1}{4}e^2B^2r^2\sin^2\theta
+e\hbar B(\frac{1}{2}+\cos\theta)\sigma_{\rho}\\
&+\frac{\hbar^2} {4r^2\sin^2\theta}+e\phi_e]\}\chi_s^{\prime},
\end{split}
\end{equation}
where the wave function $\chi_s^{\prime}$ consists of two components:
\begin{equation}\label{Wavefunction 3}
\chi_s^{\prime}=
\left (
\begin{array}{c}
\cos\frac{\theta}{2}\chi_s^+-\sin\frac{\theta}{2}\chi_s^-\\
\sin\frac{\theta}{2}\chi_s^++\cos\frac{\theta}{2}\chi_s^-
\end{array}
\right ).
\end{equation}
For the spherical surface, the geometric potential $V_g=0$. It is worth noticing that the new term $-\frac{\hbar^2}{8mr^2\sin^2\theta}$ is given by the derivative $\partial_{\phi}^2$ acting on $e^{-i\varphi}$, which connects the spin with the surface. We call this new term the spin connection geometric potential. It is clear that there are two additive terms $\frac{i\hbar^2\sigma_{\rho}}{2mr^2sin^2\theta}\partial_{\phi}$ and $\frac{e\hbar B}{4m}$, too. With $\partial_{\theta}^2$ and $\partial_{\theta}$ acting on the component $(\cos\frac{\theta}{2}\chi_s^+-\sin\frac{\theta}{2}\chi_s^-)$, we can obtain additively two spin connection geometric potentials $\frac{\hbar^2\cos\frac{\theta}{2}}{8mr^2}$ and $\frac{\hbar^2\cos\theta}{8mr^2\cos\frac{\theta}{2}}$ for $\chi_s^+$, and two other spin connection geometric potentials $-\frac{\hbar^2\sin\frac{\theta}{2}}{8mr^2}$ and $\frac{\hbar^2\cos\theta}{8mr^2\sin\frac{\theta}{2}}$ for $\chi_s^-$. Working on the other component, $(\sin\frac{\theta}{2}\chi_s^++\cos\frac{\theta}{2}\chi_s^-)$, we can get two other spin connection geometric potentials $\frac{\hbar^2\sin\frac{\theta}{2}}{8mr^2}$ and $-\frac{\hbar^2\cos\theta}{8mr^2\sin\frac{\theta}{2}}$ for $\chi_s^+$, and other two spin connection geometric potentials $\frac{\hbar^2\cos\frac{\theta}{2}}{8mr^2}$ and $\frac{\hbar^2\cos\theta}{8mr^2\cos\frac{\theta}{2}}$ for $\chi_s^-$. In the Pauli equation \eqref{Pauli Equation 8}, we notice that there is only the component $\sigma_{\rho}$ mixing with the curvature of the spherical surface and coupling to the magnetic field.\\
\indent The cylinder is a most popular geometry in the nanometer world, and its extensive investigations are various nanotubes, such as carbon nanotubes, semiconductor nanotubes, and metal nanotubes. We consider a cylinder of radius $r$, which is sketched in Fig.\ref{Cylinder}. This cylinder is described in the cylindrical coordinate system $(\theta, y, \rho)$, and is put in a magnetic field $\vec{B}$ which can be expressed as the summation of $B_0$, which is parallel to the direction of the $y$ axis, and $B_1$, which is perpendicular to the $y$ axis and parallel to the direction of the $\rho$ axis at $\theta=0$. The most suitable vector potential is $(A_{\theta},A_y,A_{\rho})=(\frac{1}{2}r^2B_0,rB_1\sin\theta,0)$, the same given by Ferrari and Cuoghi\cite{Ferrari2008}.
\indent On the cylinder, the induced Pauli matrices can be given:
\begin{equation}\nonumber
\begin{split}
&\Sigma^{\theta}=\frac{1}{r}
\left (
\begin{array}{cc}
-\sin\theta & \cos\theta\\
\cos\theta & \sin\theta
\end{array}
\right ),\\
&\Sigma^y=
\left (
\begin{array}{cc}
0 & -i\\
i & 0
\end{array}
\right ),\\
&\Sigma^{\rho}=
\left (
\begin{array}{cc}
\cos\theta & \sin\theta\\
\sin\theta & -\cos\theta
\end{array}
\right ).
\end{split}
\end{equation}
\indent On the cylinder, for the transformation of the induced Pauli matrices, the spinor rotation matrix is
\begin{equation}\nonumber
\begin{split}
\mathcal{U}(\theta,y)&=e^{\frac{i}{2}\theta\sigma_2}\\
&=
\left (
\begin{array}{cc}
\cos\frac{\theta}{2} & \sin\frac{\theta}{2}\\
-\sin\frac{\theta}{2} & \cos\frac{\theta}{2}
\end{array}
\right ),
\end{split}
\end{equation}
and its inverse matrix is
\begin{equation}\nonumber
\mathcal{U}^{-1}(\theta,y)=
\left (
\begin{array}{cc}
\cos\frac{\theta}{2} & -\sin\frac{\theta}{2}\\
\sin\frac{\theta}{2} & \cos\frac{\theta}{2}
\end{array}
\right ).
\end{equation}
The transformed induced Pauli matrices on the cylinder are
\begin{equation}\nonumber
\sigma^{\theta}=\frac{1}{r}\sigma^1,\quad
\sigma^y=\sigma^2,\quad
\sigma^{\rho}=\sigma^3.
\end{equation}
From Eq. \eqref{Pauli Equation 7}, the cylindrical surface dynamics can be described by
\begin{equation}\label{Pauli Equation 9}
\begin{split}
i\hbar\partial_t\chi_s^{\prime}=&-\frac{1}{2m}[\frac{\hbar^2}{r^2}\partial_{\theta}^2 +ie\hbar B_0\partial_{\theta}+\hbar^2\partial_y^2\\
&+2ie\hbar rB_1\sin\theta\partial_y
-(\frac{1}{4}e^2r^2B_0^2\\
&+e^2r^2B_1^2\sin^2\theta+e\hbar B_1\cos\theta\sigma_{\rho})-\frac{\hbar^2}{4r^2}]\chi_s^{\prime},\\
\end{split}
\end{equation}
where the wave function $\chi_s^{\prime}$ also has two components,
\begin{equation}\label{Wavefunction 4}
\chi_s^{\prime}=
\left (
\begin{array}{cc}
\cos\frac{\theta}{2}\chi_s^+-\sin\frac{\theta}{2}\chi_s^-\\
\sin\frac{\theta}{2}\chi_s^++\cos\frac{\theta}{2}\chi_s^-
\end{array}
\right ).
\end{equation}
On the cylinder, the geometric potential $V_g=\frac{\hbar^2}{8mr^2}$. The derivative $\partial_y^2$ cannot bring out anything from $e^{-i\varphi}$, but the derivative $\partial_{\theta}^2$ still can take out some exciting terms for the present system from $e^{-i\varphi}$. It is very interesting to us that some spin connection geometric potentials appear: $\frac{\hbar^2\cos\frac{\theta}{2}}{8mr^2}$, $\frac{\hbar^2\sin\frac{\theta}{2}}{8mr^2}$ for $\chi_s^+$, and $-\frac{\hbar^2\sin\frac{\theta}{2}}{8mr^2}$, $\frac{\hbar^2\cos\frac{\theta}{2}}{8mr^2}$ for $\chi_s^-$. Of course, the derivative $\partial_{\theta}$ can offer $\frac{ie\hbar B_0\sin\frac{\theta}{2}}{4m}$ and $-\frac{ie\hbar B_0\cos\frac{\theta}{2}}{4m}$ for $\chi_s^+$, and $\frac{ie\hbar B_0\cos\frac{\theta}{2}}{4m}$ and $\frac{ie\hbar B_0\sin\frac{\theta}{2}}{4m}$ for $\chi_s^-$. In the Pauli equation \eqref{Pauli Equation 9}, there is only the component $\sigma_{\rho}$ coupling to the magnetic field with $-\frac{1}{2m}e\hbar B_1\cos\theta\sigma_{\rho}$.\\
\indent The torus is a mathematical topological geometry. It is very interesting to investigate, both from the theoretical and from the experimental points of view. In the reference system $(\theta,\phi,\rho)$, a torus is put in an arbitrary constant magnetic field $\vec{B}$, which is sketched in Fig. \ref{Torus}.
\indent On the torus, the induced Pauli matrices are
\begin{equation}\nonumber
\begin{split}
&\Sigma^{\theta}=\frac{1}{r}
\left (
\begin{array}{cc}
-\sin\theta & \cos\theta e^{-i\phi}\\
\cos\theta e^{i\phi} & \sin\theta
\end{array}
\right ),\\
&\Sigma^{\phi}=\frac{1}{R}
\left (
\begin{array}{cc}
0 & -ie^{-i\phi}\\
ie^{i\phi} & 0
\end{array}
\right ),\\
&\Sigma^{\rho}=
\left (
\begin{array}{cc}
\cos\theta & \sin\theta e^{-i\phi}\\
\sin\theta e^{i\phi} & -\cos\theta
\end{array}
\right ).
\end{split}
\end{equation}
The spinor rotation matrix is
\begin{equation}\nonumber
\begin{split}
\mathcal{U}(\theta,\phi)&=e^{\frac{i}{2}\theta\sigma_2+\frac{i}{2}\phi\sigma_3}\\
&=
\left (
\begin{array}{cc}
\cos\frac{\theta}{2}e^{i\frac{\phi}{2}} & \sin\frac{\theta}{2} e^{-i\frac{\phi}{2}}\\
-\sin\frac{\theta}{2}e^{i\frac{\phi}{2}} & \cos\frac{\theta}{2}e^{-i\frac{\phi}{2}}
\end{array}
\right ),
\end{split}
\end{equation}
and its inverse matrix is
\begin{equation}\nonumber
\mathcal{U}^{-1}(\theta,\phi)=
\left (
\begin{array}{cc}
\cos\frac{\theta}{2}e^{-i\frac{\phi}{2}} & -\sin\frac{\theta}{2}e^{-i\frac{\phi}{2}}\\
\sin\frac{\theta}{2}e^{i\frac{\phi}{2}} & \cos\frac{\theta}{2}e^{i\frac{\phi}{2}}
\end{array}
\right ).
\end{equation}
The transformed induced Pauli matrices on the torus are
\begin{equation}\nonumber
\sigma^{\theta}=\frac{1}{r}\sigma^1,\quad \sigma^{\phi}=\frac{1}{R}\sigma^2,\quad \sigma^{\rho}=\sigma^3.
\end{equation}
According to the gauge condition \eqref{Gauge Choice 1}, the most suitable $(A_{\theta},A_{\phi},A_{\rho})=(\frac{1}{2}B_1r\sin\phi(R_0\sin\theta+r), \frac{1}{2}R(B_0R-B_1r\cos\theta\cos\phi),0)$ with $R=R_0+r\sin\theta$. From Eq. \eqref{Pauli Equation 7}, we can calculate the surface Pauli equation:
\begin{equation}\label{Pauli Equation 10}
\begin{split}
i\hbar\partial_t\chi_s^{\prime}=&-\frac{1}{2m}\{\frac{\hbar^2}{r^2} \partial_{\theta}^2+\frac{\hbar^2\cos\theta}{rR}\partial_{\theta} +\frac{\hbar^2}{R^2}\partial_{\phi}^2\\
&-\frac{\hbar^2}{4R^2}-\frac{i\hbar^2}{R^2}\sigma_{\rho}\partial_{\phi}\\
&+\frac{ie\hbar}{r}B_1(R_0\sin\theta+r)\sin\phi\partial_{\theta}\\
&+\frac{ie\hbar}{R}(B_0R-B_1r\cos\theta\cos\phi) \partial_{\phi}\\
&+\frac{e\hbar}{2R}(B_0R-B_1r\cos\theta\cos\phi)\sigma_{\rho}\\
&+\frac{R_0^2+r^2+ 2rR_0\sin\theta}{2rR}ie\hbar B_1\cos\theta\sin\phi\\
&+\frac{r}{2R}ie\hbar B_1\cos\theta\sin\phi\\
&-\frac{e^2}{4}[B_1^2\sin^2\phi(R_0\sin\theta+r)^2\\ &+(B_0R-B_1r\cos\theta\cos\phi)^2]\\
&-e\hbar(B_0+\frac{r}{R}B_1\cos\theta\cos\phi)\cos\theta\sigma_{\rho}\\
&+\frac{\hbar R_0^2}{(2rR)^2}\} \chi_s^{\prime},
\end{split}
\end{equation}
where the transformed surface wave function $\chi_s^{\prime}$ can be expressed as
\begin{equation}
\chi_s^{\prime}=
\left (
\begin{array}{c}
\cos\frac{\theta}{2}\chi_s^+-\sin\frac{\theta}{2}\chi_s^-\\
\sin\frac{\theta}{2}\chi_s^++\cos\frac{\theta}{2}\chi_s^-
\end{array}
\right ).
\end{equation}
It is very interesting to notice that there is $\frac{\hbar^2}{8mR^2}$, the fourth term on the right-hand side of Eq. \eqref{Pauli Equation 10}, which is a new spin connection geometric potential, and plays the same role as the well-known geometric potential $V_g=-\frac{\hbar^2R_0^2}{8mr^2R^2}$, but both terms have opposite sign. In other words, $V_g$ is attractive, but $\frac{\hbar^2}{8mR^2}$ is repulsive. In the present system, the derivatives $\partial_{\theta}^2$ and $\partial_{\theta}$ can contribute some new spin connection geometric potentials $\frac{\hbar^2\cos\frac{\theta}{2}}{8mr^2}$, $\frac{\hbar^2\sin\frac{\theta}{2}}{8mr^2}$, $\frac{\hbar^2\cos\theta\sin\frac{\theta}{2}}{4mrR}$, and $-\frac{\hbar^2\cos\theta\cos\frac{\theta}{2}}{4mrR}$ for $\chi_s^+$, and $-\frac{\hbar^2\sin\frac{\theta}{2}}{8mr^2}$, $\frac{\hbar^2\cos\frac{\theta}{2}}{8mr^2}$, $\frac{\hbar^2\cos\theta\cos\frac{\theta}{2}}{4mrR}$, and $\frac{\hbar^2\cos\theta\sin\frac{\theta}{2}}{4mrR}$ for $\chi_s^-$. In the Pauli equation \eqref{Pauli Equation 10}, there is only the component $\sigma_{\rho}$ mixing with the curvature of the toroidal surface and coupling to the magnetic field.
\section{Conclusions}
\indent In this paper we have briefly reviewed, reconsidered, and extended the thin-layer quantization scheme, which was used to derive the surface Schr\"{o}dinger equation for a spinless particle constrained on a spatially curved surface, to derive the surface Pauli equation for a charged spin particle confined on a curved surface in an electromagnetic field. Through a rotation, we have described the reduced system in the transformed spinor representation. In this case, we have chosen the proper gauge condition for the electromagnetic field to accomplish analytically the separation of the Pauli equation into normal and surface parts. For the separation of the Schr\"{o}dinger equation for a reduced system, the fundamental evidences \cite{Ferrari2008} are still valid for the decoupling of the Pauli equation for the constrained system. There is no coupling between the electromagnetic field and the surface, and the separation of normal and surface parts can be accomplished analytically with proper gauge choice. It is worth noticing that the two fundamental evidences are valid with the metric tensor $\tilde{G}_{ij}$. In other words, the thin-layer quantization scheme is valid to systems without source current perpendicular to $\mathcal{S}$.\\
\indent In order to understand the connection between spin and the surface and the geometric potential clearly, we have applied the surface Pauli equation to spherical, cylindrical, and toroidal surfaces. The tangent variable derivatives act on the rotation $\mathcal{U}^{-1}(q_1,q_2)$ to contribute some spin connection geometric potentials for the three examples. These spin connection geometric potentials are produced in the inverse process of the transformation for the spinor representation. The results support that the curved surface acts on a spin particle with more contributions. In the special case, for the three examples there is only the component $\sigma_{\rho}$ appearing in the surface Pauli equations.
\section{Acknowledgment}
\indent This work is supported by the National Natural Science Foundation of China (under Grant No. 11047020, No. 11404157, No. 11274166, No. 11275097, and No. 11475085), the National Basic Research Program of China (under Grant No. 2012CB921504) and the Natural Science Foundation of Shandong Province of China (under Grant No. ZR2012AM022, and No. ZR2011AM019).

\end{document}